%% file: paper_arx.tex
\documentclass[aps,prl,showpacs,twocolumn,groupedaddress]{revtex4}  
\usepackage{graphicx}  
\usepackage{dcolumn}   
\usepackage{bm}        
\usepackage{amssymb}   

\newcommand{\dmes}       {\ensuremath{ D}}
\newcommand{\ad}       {\ensuremath{\bar{ D}}}

\newcommand{\adzero}     {\ensuremath{\bar{ D}^0}}
\newcommand{\dst}       {\ensuremath{ D^{*}}}
\newcommand{\adst}       {\ensuremath{\bar{ D}^{*}}}

\newcommand{\adstzero}   {\ensuremath{\bar{ D}^{*0}}}

\newcommand{\dstminus}  {\ensuremath{ D^{*-}}}

\newcommand{\adststzero} {\ensuremath{\bar{ D}^{**0}}}
\newcommand{\dstst}     {\ensuremath{ D^{**}}}
\newcommand{\adstst}     {\ensuremath{\bar{ D}^{**}}}

\newcommand{\effpi}      {\ensuremath{\varepsilon_\pi}}

\begin{document}

\hspace{5.2in} \mbox{FERMILAB-Pub-04/284-E}

\title{Measurement of the Ratio of $B^+$ and $B^0$ Meson Lifetimes}
\input list_of_authors_r2.tex  
\date{\today}

\begin{abstract}
The ratio of $B^+$ and $B^0$ meson lifetimes
was measured
using data collected in 2002--2004 by the D\O\ experiment in Run II of 
the Fermilab Tevatron Collider. 
These mesons were reconstructed in
$B \to \mu^+ \nu \dstminus X$ decays, which are
dominated by $B^0$, and $B \to \mu^+
\nu \adzero X$ decays, which are dominated by $B^+$.  The ratio of
lifetimes
is measured to be
$\tau^+/\tau^0 = 1.080 \pm 0.016\ \mbox{(stat)} \pm 0.014\ \mbox{(syst)}$.
\end{abstract}
\maketitle


In the last few years, significant progress has been made, on both
experimental and theoretical fronts, in the understanding of the 
lifetimes of hadrons containing heavy quarks. Charm and bottom meson
(except $B_c$) lifetimes have been measured with precisions ranging 
from 0.5\% to 4\%, although lifetimes of heavy baryons 
are not known as well \cite{pdg}. On the theoretical front, 
predictions are being 
made using a rigorous approach based on the heavy quark expansion 
(in negative powers of the heavy quark mass) \cite{bigi},
where the large mass of the bottom quark
considerably simplifies calculations.
Theoretical uncertainties are further reduced for ratios of
lifetimes. For instance, the ratio of the
$B^+$ and $B^0$ lifetimes has been predicted to be $1.06 \pm 0.02$
\cite{tarantino}.
Experimentally, ratios of lifetimes have smaller uncertainties, since
many common sources of systematics cancel.

%

In this Letter, we present a measurement of the ratio of 
$B^+$ and $B^0$ lifetimes 
using a large sample of semileptonic
$B$ decays collected by the D\O\ experiment at Fermilab in $p \bar{p}$
collisions at $\sqrt{s}=1.96$ TeV.  The data correspond
to approximately 440 pb$^{-1}$ of integrated luminosity.
$B$ mesons were selected via their decays $B \to \mu^+ \nu \adzero X$ 
\cite{charge}
and were classified into two exclusive groups: a ``\dstminus'' sample,
containing all events with reconstructed $\dstminus \to \adzero \pi^-$ decays,
and a ``\adzero'' sample, containing all remaining events. Both simulation
and available experimental results show that the \dstminus\ sample is dominated
by $B^0 \to \mu^+ \nu \dstminus X$ decays, while the \adzero\ sample is
dominated by $B^+ \to \mu^+ \nu \adzero X$ decays. 

The classification into these two samples was based on the presence of
a slow pion from $\dstminus \to \adzero \pi^-$ decay, and thus
was independent of the $B$-meson lifetime. Therefore, the ratio of
the number of events in the two samples, expressed as a function of
the proper decay length, depends mainly on the lifetime difference
between the $B^+$ and $B^0$ mesons. The influences of the
selection criteria, detector properties, and some systematic
uncertainties are significantly reduced.

The D\O\ detector is described in detail elsewhere \cite{dzero}. The
detector components most important to this analysis are the
central tracking and muon systems. The tracking system
consists of a silicon microstrip tracker and a central fiber
tracker, both located within a 2\ T superconducting solenoidal
magnet.
The resolution for the distance of closest approach as provided
by the tracking system is $\approx 50~\mu$m
for tracks with $p_T \approx 1$ GeV/$c$, improving asymptotically
to 15 $\mu$m for tracks with $p_T \ge$~10 GeV/$c$, where
$p_T$ is the component
of the momentum perpendicular to the beam pipe.
The muon system is located outside the calorimeters
and consists of a layer of drift chambers and scintillation trigger counters
in front of 1.8~T toroids, followed by two more similar layers after
the toroids. 

Events with semi-muonic $b$-hadron decays were selected using a suite of
inclusive single-muon triggers in a three-level trigger
system. 
Muons were identified by extrapolating tracks found in the central
tracking system and matching them with muon track segments formed from
hits in the muon system. Muons were required to have a transverse
momentum $p_T^{\mu} > 2$ GeV/$c$ and total momentum $p^{\mu} > 3$
GeV/$c$.

The primary vertex of the $p \bar{p}$ interaction was determined for
each event.
The average
position of the beam-collision point was included as a constraint.  The
precision of the primary vertex reconstruction was on average about 20
$\mu$m in the plane perpendicular to the beam direction
and about 40 $\mu$m along the beam direction.


\adzero\ candidates were selected using their $\adzero \to K^+ \pi^-$
decay mode.  All charged particles in an event were clustered into
jets using the DURHAM clustering algorithm \cite{dur} with a jet $p_T$
cut-off parameter of 15 GeV/$c$ \cite{pyth}.  The \adzero\ candidate was
constructed from two particles of opposite charge belonging to the
same jet as the reconstructed muon. Both particles were required to
have $p_T > 0.7$ GeV/$c$ and to form a common \adzero\ vertex. The $p_T$
of the \adzero\ was required to exceed 5 GeV/$c$. To reduce combinatorial
background, we required the \adzero\ vertex to have a positive displacement
in the $xy$ plane, relative to the primary vertex, with at least $4\sigma$
significance. Although this last requirement can bias the lifetime
distribution of a $B$ candidate, our analysis procedure of determining
the ratio of $B^+$ and $B^0$ events in bins of proper time should remove this
bias in the final result.  The trajectory of the muon and \adzero\ candidates 
were required to originate from a common $B$ vertex. The
$\mu^+ \adzero$ system was required to have an invariant mass between 2.3
and 5.2 GeV/$c^2$.

\begin{figure}
\vspace*{-0.7cm}
\hspace*{-0.3cm}
\mbox{
\includegraphics[scale=0.45]{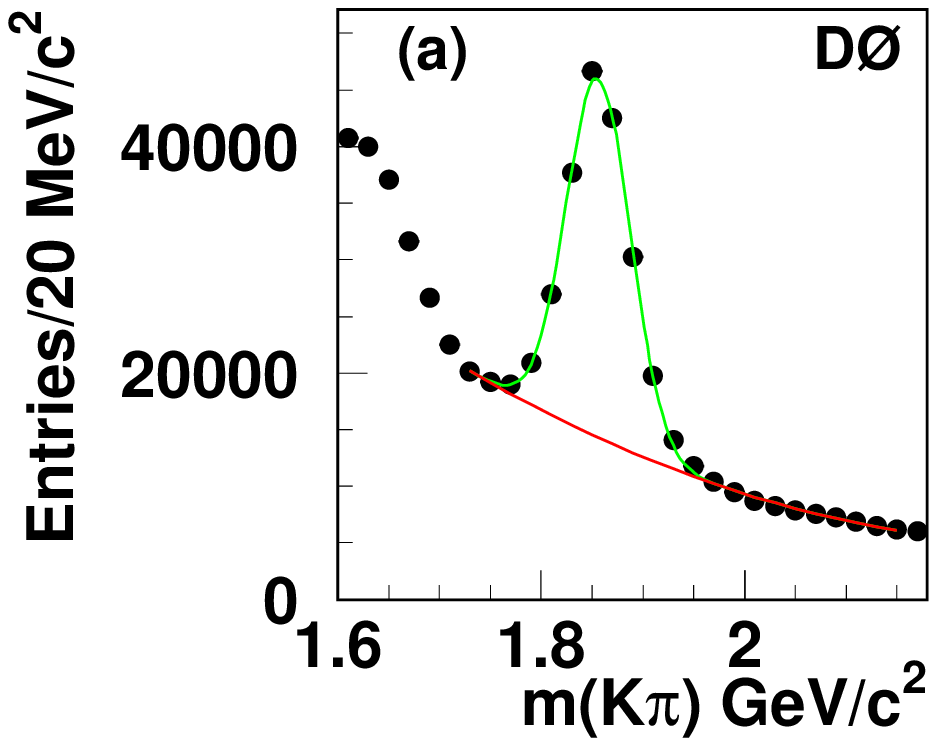}
\includegraphics[scale=0.45]{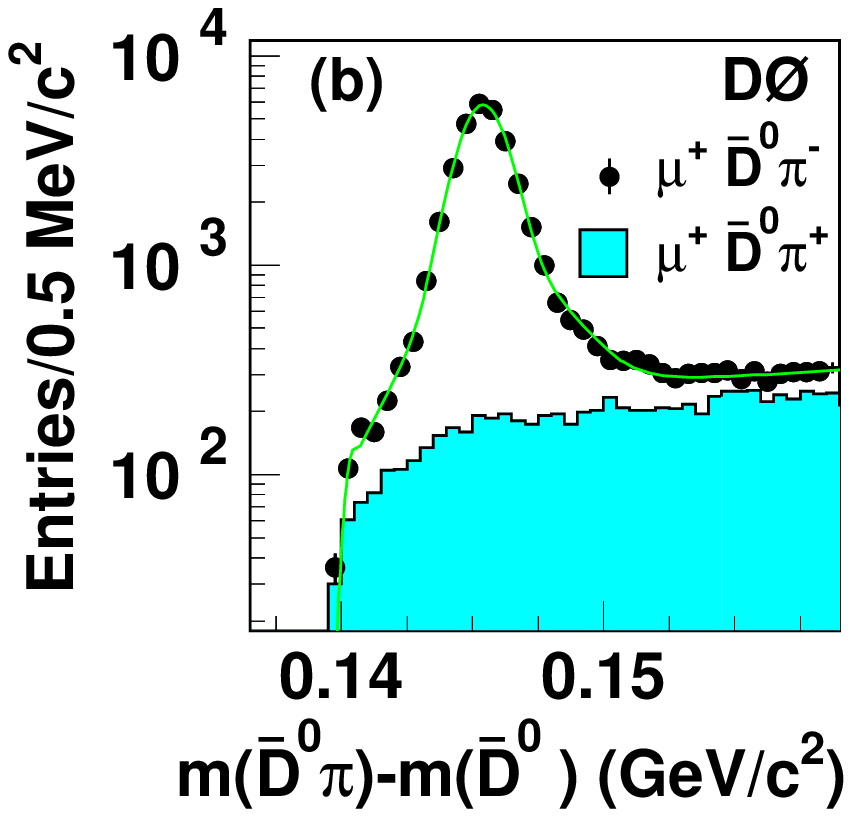}
}
\caption{\label{fig1}
    (a)\ Invariant mass of the $K \pi$ system.
    The curve shows the result of the fit of the $K^+ \pi^-$ mass 
    distribution with the sum of a signal Gaussian function 
    and polynomial background function.
    (b)\ Mass difference $\Delta m = m(\adzero \pi) - m(\adzero)$.
}
\end{figure}

The masses of the kaon and pion were assigned to the two tracks according 
to the charge of the muon, assuming the $\mu^+ K^+ \pi^-$ 
combination. The mass spectrum of the $K \pi$ 
system after these selections is shown in Fig.\ \ref{fig1}(a). The signal
in the \adzero\ peak contains $126073 \pm 610$ events.

All reconstructed $\mu^+ \adzero$ events were classified into three
non-overlapping samples. For each $\mu^+ \adzero$ candidate, a search
was made for an additional pion.
The mass difference $\Delta m = m(\adzero \pi) -
m(\adzero)$ for all such pions, when $1.8 < m(\adzero) < 1.9$
GeV/$c^2$, is shown in Fig.\ \ref{fig1}(b). The peak in this figure
corresponds to the production of the $\mu^+ \dstminus$ system. All
events containing a pion with a charge opposite to that of the muon 
(right-charge combination) and $0.1425 < \Delta m < 0.1490$ GeV/$c^2$ were
included in the \dstminus$^R$ sample. All events containing a pion with
the same charge as the muon (wrong-charge combination) and $0.1425 <
\Delta m < 0.1490$ GeV/$c^2$ were included in the auxiliary \dstminus$^W$
sample.  This sample contains true \adzero\ but fake \dstminus\ events 
and gives an estimate of the combinatorial
background for selected $\mu^+ \dstminus$ candidates. The $\Delta m$
distribution for such events is shown in Fig.\ \ref{fig1}(b) as the filled
histogram.  All remaining events were assigned to the \adzero\ sample.


Since the final (semileptonic) state has missing particles, including
the neutrino, the proper
decay length was not determined. Instead, for each reconstructed
candidate, the measured visible proper decay length $x^M$
was computed as $x^M = m_B c \left( \mbox{\boldmath $L$}_T \cdot
\mbox{\boldmath $p$}_T(\mu^+ \adzero) \right) / 
|\mbox{\boldmath $p$}_T(\mu^+ \adzero)|^2$. 
{\boldmath $L$}$_T$
is the vector in the axial plane from the primary to the $B$-meson decay
vertex, {\boldmath $p$}$_T(\mu^+ \adzero)$ is the transverse momentum of
the $\mu^+ \adzero$ system and $m_B$ is the mass of the $B$ meson, for
which the value 5.279 GeV/$c^2$ was used \cite{pdg}.
The pion from the \dstminus\ decay was not used for the computation 
of the transverse momentum and the decay length.

Candidates in each of the samples were divided into eight groups according to
their $x^M$ value. The number of $\mu^+ \adzero$ events $N^{*R}_i$
(from the \dstminus$^R$ sample), $N^{*W}_i$ (from the \dstminus$^W$ sample), 
and $N^0_i$
(from the \adzero\  sample) in each interval $i$ (where $i$ ranges from
one to eight) were determined from the 
fit of the $K\pi$ mass spectrum between 1.72 and 2.16 GeV/$c^2$ 
with the sum of a Gaussian signal
function and a polynomial background function. 
The mean and width of the Gaussian function were fixed to
the values obtained from the fit of the overall mass distribution in
each sample. The fitting procedure was the same for all samples.
Table \ref{tab1} gives the numbers obtained for each $x^M$ interval.

The number of $\mu^+ \dstminus$ events for each interval $i$ of $x^M$
was defined as $N_i(\mu^+ \dstminus) = N^{*R}_i - C \cdot N^{*W}_i$,
where $C \cdot N^{*W}_i$ accounts for the combinatorial background under the
$\dstminus$ peak as shown in Fig.\ \ref{fig1}(b). The coefficient $C = 1.27 \pm 0.03$
reflects the difference in the combinatorial background between
$\mu^+ \adzero \pi^-$ and $\mu^+ \adzero \pi^+$ events. It was
determined from the ratio of the numbers of these events in the interval 
$0.153 < \Delta m <
0.160$ GeV/$c^2$. The number of $\mu^+ \adzero$ events in each
interval $i$ in $x^M$ was defined as $N_i(\mu^+ \adzero) = N^0_i +
N^{*W}_i + C \cdot N^{*W}_i$.


The experimental observable $r_i$ is the ratio of $\mu^+ \dstminus$ and $\mu^+
\adzero$ events in interval $i$ of $x^M$, {\rm i.e.,} $r_i = N_i(\mu^+
\dstminus)/N_i(\mu^+ \adzero)$. Values of $r_i$ and statistical uncertainties
 are given in Table\ \ref{tab1}.  
The measurement of the lifetime difference between $B^+$ and $B^0$ is
given by $k \equiv \tau^+/\tau^0 -1$. It was determined from the
minimization of $\chi^2(\effpi,k)$:
\begin{equation}
\chi^2(\effpi,k) = \sum_i \frac{(r_i - r_i^e(\effpi,k))^2}{\sigma^2(r_i)},
\label{chi2}
\end{equation}
where $r_i^e(\effpi,k)$ is the expected ratio of $\mu^+ \dstminus$
and $\mu^+ \adzero$ events, and \effpi\ is the efficiency to
reconstruct the slow pion in the $\dstminus \to \adzero \pi^-$ decay.
$\effpi$ was assumed to be independent of $x^M$ and, along with $k$, was a
free parameter in the minimization. We present evidence for the
validity of this assumption in the discussion of systematic uncertainties.
The sum $\sum_i$ was taken over all intervals with positive $x^M$.

\begin{table*}
\caption{\label{tab1} Definition of the intervals in visible proper decay 
length, $x^M$. For each interval $i$, the number of events in 
the \dstminus$^R$, 
\dstminus$^W$ and \adzero\ samples, the ratio $r_i$, and the expected 
value $r_i^e$ for $\tau^+/\tau^0-1 = 0.080$ are given.}
\begin{ruledtabular}
\newcolumntype{A}{D{A}{\pm}{-1}}
\newcolumntype{B}{D{B}{-}{-1}}
\begin{tabular}{cBAAAAc}
$i$ & 
\multicolumn{1}{c}{$x^M$ range (cm)}  & 
\multicolumn{1}{c}{$N_i^{*R}$} & 
\multicolumn{1}{c}{$N_i^{*W}$} & 
\multicolumn{1}{c}{$N_i^0$} & 
\multicolumn{1}{c}{$r_i$} & 
	$r_i^e$ \\ \hline
1 & -0.1\ B \ 0.0 & 1714 \ A \  53 & 89 \ A \  22 & 5225 \ A \  151 & 
  0.295\ A \ 0.015  & 0.309 \\
2 & 0.0\ B \ 0.02   & 6213 \ A \  94 & 200 \ A \  28 & 18134 \ A \  222 & 
  0.321\ A \ 0.007  & 0.315 \\
3 & 0.02\ B \ 0.04  & 5941 \ A \  91 & 169 \ A \  22 & 17703 \ A \  208 & 
  0.317\ A \ 0.007  & 0.313 \\
4 & 0.04\ B \ 0.07  & 6424 \ A \  94 & 213 \ A \  23 & 19707 \ A \  216 &
  0.305\ A \ 0.006  & 0.308 \\
5 & 0.07\ B \ 0.10  & 4029 \ A \  74 & 115 \ A \  17 & 12885 \ A \  171 & 
  0.295\ A \ 0.007  & 0.300 \\
6 & 0.10\ B \ 0.15  & 3459 \ A \  68 & 106 \ A \  16 & 11532 \ A \  162 &
  0.282\ A \ 0.007  & 0.291 \\
7 & 0.15\ B \ 0.25  & 2253 \ A \  57 & 58 \ A \  13  & 7567 \ A \  137 & 
  0.283\ A \ 0.009  & 0.276 \\
8 & 0.25\ B \ 0.40  & 518 \ A \  28  & 2 \ A \  6    & 1875 \ A \  75 & 
  0.274\ A \ 0.019  & 0.256
\end{tabular}
\end{ruledtabular}
\end{table*}


Information used to determine the expected ratio, $r_i^e(\effpi,k)$,
included both experimental measurements as well as results from Monte
Carlo simulations.  For the $j$th $B$-meson decay channel, the
distribution of the visible proper decay length ($x$) is given by 
$P_j(x) = \int dK\  D_j(K)
\cdot \theta(x) \cdot \frac{K}{c\tau_j} \exp(-\frac{Kx}{c\tau_j})$.
%
$\tau_j$ is the lifetime of the $B$ meson, the $K$-factor, $K =
p_T^{\mu^+ \adzero}/p_T^B$, reflects the difference between the observed
and true momentum of the $B$ meson, and $\theta(x)$ is the step function.
The function $D_j(K)$ is the normalized distribution of the $K$-factor for
the $j$th decay channel.

Transformation from the true value of $x$ to the experimentally 
measured value $x^M$ 
is given by $f_j(x^M) = \int dx\  R_j(x-x^M) \cdot \varepsilon_j(x) 
\cdot P_j(x)$,
where $R_j(x-x^M)$ is the detector resolution, and 
$\varepsilon_j(x)$ is the reconstruction efficiency of $\mu^+ \adzero$ for
the $j$th decay. It does not include \effpi\ for channels with \dstminus.
Finally, the expected value $r_i^e(\effpi,k)$  
is given by:
\begin{equation}
r_i^e(\effpi,k)  = \frac{\effpi \cdot F^*_i(k)}
	                {F^0_i(k) + (1-\effpi) \cdot F^*_i(k)}.
\end{equation}
Here $F^{*,0}_i = \int_i dx^M \sum_j Br_j \cdot f_j(x^M)$
with the summation $\sum_j$ taken over all decays 
to \dstminus\ (\adzero) for $F^*_i$ ($F^0_i$). 

For the computation of $r_i^e$, the world average of the $B^+$ lifetime 
\cite{pdg} was used. The $B^0$ 
lifetime $\tau^0$ was expressed as $\tau^0 = \tau^+/(1+k)$.
The branching fractions 
$B \to \mu^+ \nu \ad$ and $B \to \mu^+ \nu \adst$ were taken from Ref. 
\cite{pdg}.
The following branching fractions were derived from experimental measurements 
\cite{pdg,aleph,delphi,comb}:
$Br(B^+ \to \mu^+ \nu \adststzero) =  (2.67 \pm 0.37) \% $,
$Br(B^+ \to \mu^+ \nu \adststzero  \to \dstminus X) 
  =  (1.07 \pm 0.25)\%$, and 
$Br(B^0_s \to \mu^+ \nu \dmes_s^{**-}) = (2.3^{+2.4}_{-2.3})\%$.
$\dstst$ states include both narrow and wide $\dstst$ resonances and
non-resonant $\dmes X$ and $\dst X$ production.
Regarding the possible decays of $\dmes_s^{**-}$, there is no 
experimental data on the $Br(\dmes_s^{**-} \to \dstminus X)$. 
Its central value was therefore set to 0.35 and it was varied 
between 0.0 and 1.0 to estimate the systematic uncertainty from this source.
All other branching fractions were derived assuming isospin invariance.


The distributions $D_j(K)$, $R_j(x)$,
and $\varepsilon_j(x)$ were taken from the Monte Carlo simulation, and the 
corresponding systematic uncertainties were taken into account.
All processes involving $b$ hadrons were simulated using the 
{\sc EvtGen} \cite{EvtGen} generator interfaced 
to {\sc pythia} \cite{pyth} and followed 
by the full modeling of the detector response and event reconstruction.
The semileptonic $b$-hadron decays were generated using 
the ISGW2 model \cite{isgw2}.


Assuming the given branching fractions and reconstruction efficiencies, 
the decay $B \to \mu^+ \dstminus X$ contains $(89\pm 3)$\% $B^0$, 
$(10\pm 3)$\% $B^+$, and $(1\pm 1)$\% $B_s^0$, while the 
decay $B \to \mu^+ \adzero X$ contains 
$(83\pm 3)$\% $B^+$, $(15\pm 4)$\% $B^0$, and $(2\pm 1)$\% $B_s^0$.

A special study showed that in addition to the main decay
$B \to \mu^+ \adzero X$, the decay 
$B \to \tau^+ \adzero X \to \mu^+ \nu \bar{\nu} \adzero X$ results in a 
($5 \pm 2$)\% contribution and the process $c \bar{c} \to \mu^+ \adzero X$ 
a ($10 \pm 7$)\%
contribution to the selected $\mu^+ \adzero$ sample. 
These processes
were taken into account in the analysis. 

\begin{figure}
\vspace*{-0.7cm}
    \includegraphics[scale=0.47]{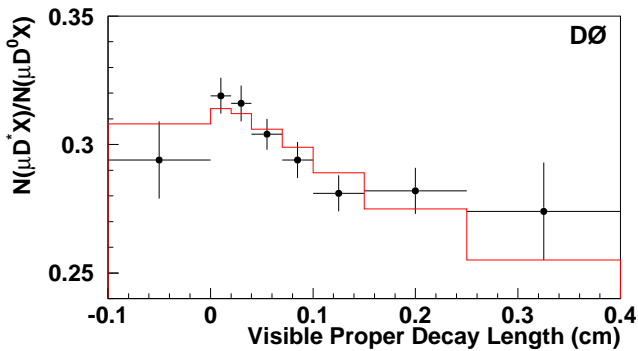}
    \caption{Points with the error bars show the ratio of the number 
       of events in the 
       $\mu^+ \dstminus$ and $\mu^+ \adzero$ samples
       as a function of the visible proper decay length. The result
       of the minimization of Eq.\ (\ref{chi2}) with $k=0.080$ is shown
       as a histogram.}
    \label{fig7}  
\end{figure}

Using all these inputs, the minimization of the
$\chi^2$ distribution, Eq.\ (\ref{chi2}), gives:
$k \equiv  \tau^+/\tau^0 - 1  =  0.080 \pm 0.016 \ \mbox{(stat)}$.
The $\chi^2$ at the minimum is 4.2 for 5 d.o.f, \effpi\ is 
$0.864 \pm 0.006\ \mbox{(stat)}$, and the global correlation 
coefficient between $k$ and \effpi\ is 0.18. The simulation predicted 
$\effpi = 0.877 \pm 0.003$.  The reasonable agreement in \effpi\ between data
and simulation reflects good consistency 
of input efficiencies and branching fractions with 
experimental data. Figure \ref{fig7} presents
the $r_i$ values together with the result of the fit.

The influence of various sources of systematic uncertainty on the final
result is summarized in Table \ref{tab3}. Different contributions can be 
divided into three groups. The first part includes 
uncertainties coming from 
the experimental measurements,
{\rm e.g.}, branching fractions and lifetimes. All inputs were varied 
by one standard deviation. Only the most significant contributions 
are listed as individual entries in Table \ref{tab3}; all remaining
uncertainties are combined into a single entry ``other contributions.'' 

The second group includes uncertainties due to the inputs taken from
the Monte Carlo simulation. They were estimated as follows.
The uncertainty due to the decay length dependence of the efficiencies 
$\varepsilon(B \to \mu^+ \nu \adzero X)$ was obtained
by repeating the analysis with decay length independent efficiencies 
used for all decay modes. This dependence almost cancels in the ratio 
of the number of events in the two samples, 
leading to the reduced systematic uncertainty in $\tau^+/\tau^0$.


The variation of the efficiency from channel to channel arises from 
differences in the kinematics of $B$-meson decays and thus depends on their 
modeling in simulation. 
To estimate the uncertainty in the efficiency due to this effect,
an alternative HQET model of $B$-meson decays \cite{hqet}
was implemented, and 
the selection cuts
on the $p_T$ of the $\mu^+$ and \adzero\ were varied over a wide range.

The same alternative model and the variation of $p_T$ cuts were used 
to study the model dependence of the $K$-factors. 
In all cases, 
the variation of the average value of $K$-factors 
did not exceed 2\%. 
Distributions of $K$-factors were determined separately for
$B \to \mu^+ \nu \adzero$, $B \to \mu^+ \nu \adst$, 
$B \to \mu^+ \nu \adstst \to \adzero X$, and 
$B \to \mu^+ \nu \adstst \to \adst X$. To estimate the uncertainty
due to the modeling of $\adstst$ decays, which include
both resonant and non-resonant components and are not yet well understood, 
the analysis was repeated with the distributions of $K$-factors from 
$B \to \adstst \to \adzero (\adst)$ decays 
set to be the same as for $B \to \adzero (\adst)$ decays.


The selection of the slow pion was made independently of the $B$ lifetime,
and the efficiency \effpi\ was assumed constant in the minimization. 
A dedicated study of $K_S \to \pi^+ \pi^-$ decays 
showed good stability of the track reconstruction efficiency 
with the change of decay length over a wide range.
The slope in the efficiency was estimated to be
$0.0038 \pm 0.0059$ cm$^{-1}$.
The independence of \effpi\ on the decay length was also verified 
in simulation, where no deviation from the constant value was detected 
within available simulation statistics. 



The average decay length resolution, approximately 35 $\mu$m for 
this measurement,
and the fraction of events with larger resolution, modeled by a
Gaussian function with resolution of 1700 $\mu$m, were varied over a wide
range, significantly exceeding the estimated difference in resolution
between data and simulation. 


The ratio of events with negative decay length in \dstminus\ and \adzero\ 
samples (the first row in Table \ref{tab1}) is sensitive to the differences
in resolution of these two samples. The comparison of this 
ratio in data with the simulation was used to estimate the systematic
uncertainty due to this difference.

\begin{table}[t]
\caption{\label{tab3}Summary of systematic uncertainties.}
\begin{ruledtabular}
\begin{tabular}{lc}
Source & $\Delta(\tau^+/\tau^0)$ \\ \hline
$Br(B^0 \to \mu^+ \nu \dstminus)$ & 0.0005 \\
$Br(B^+ \to \mu^+ \nu \adstzero)$  & 0.0010 \\
$Br(B^+ \to \mu^+ \nu \adststzero)$  & 0.0009 \\
$Br(B^+ \to \mu^+ \nu \dstminus \pi^+ X$) & 0.0059 \\
$Br(B^0_s \to \mu^+ \nu D_s^- X$) & 0.0009 \\
$\dmes_s^{**-} \to \dstminus X$ & 0.0020 \\ 
$c\bar{c} \to \mu^+ \nu \adzero X $ contribution        &   0.0015 \\
Other contributions & 0.0006 \\
\hline
$\varepsilon(B \to \mu^+ \nu \adzero X)$, decay length dependence & 0.0014 \\
$\varepsilon(B \to \mu^+ \nu \adzero X)$, average value  & 0.0030 \\
\effpi, decay length dependence & 0.0036 \\
decay length resolution & 0.0024 \\
Difference in \dstminus\  and \adzero\  resolution & 0.0053 \\
$K$-factors, average value  & 0.0032 \\ 
$K$-factors, difference between channels  & 0.0013 \\ \hline
Fitting procedure & 0.0086 \\ 
Background level under \dstminus\  & 0.0004 \\
\hline
{\bf Total}  & {\bf 0.0136} \\ 
\end{tabular}
\end{ruledtabular}
\end{table}

Since the $\mu^+ \dstminus$ and $\mu^+ \adzero$ event yields 
as a function of proper decay length are extracted by fitting the respective 
mass distributions with signal and background functional forms, 
the fitting procedure can be another source of systematic uncertainty.  
Different parameterizations of
signal and background functions were used. The maximal variation of the
result obtained was taken as the systematic uncertainty due to this source.
Finally, the uncertainty in the background level under the \dstminus\ peak
in Fig.\ \ref{fig1}(b) was also taken into account.

Estimated systematic uncertainties from different sources were
added in quadrature 
and the total systematic uncertainty on the ratio of lifetimes is
$\sigma(\tau^+ / \tau^0) = 0.014$.
Various consistency checks of this measurement were also
performed.  The total sample of events was divided into two parts using
different criteria, such as the sign of the muon rapidity, polarity of the
solenoid, charge of the muon, $p_T$ of the muon, position of the primary
interaction, etc. The measurement was repeated independently for each
sample. The definition of proper decay length intervals was varied, 
one more interval, 0.4 -- 0.8 cm, was added, and the last interval, 
0.25 -- 0.4 cm, was removed from the fit. 
In all cases, the results are consistent
within statistical uncertainties.  Finally, the measurement of the ratio
of lifetimes was performed with simulated events. The resulting value
$k^{\rm MC} = 0.084 \pm 0.015$ agrees well with the generated lifetime
ratio $k^{\rm gen} = 0.070$.

In summary, the ratio of $B^+$ and $B^0$ meson lifetimes
was found to be:
\begin{equation}
k = \frac{\tau^+}{\tau^0} - 1 = 0.080 \pm 0.016 \ \mbox{(stat)}
\pm 0.014 \ \mbox{(syst)}.
\end{equation}
This result is the most precise measurement of this parameter,
and agrees well with the world average value $k= 0.086 \pm 0.017$ \cite{pdg}.
Improved precision of the ratio of $B^+$ and $B^0$ lifetimes will allow
a better test of theoretical predictions, especially those inputs
to the calculations that rely on lattice QCD or on other non-perturbative
methods \cite{bigi,tarantino}.

\input acknowledgement_paragraph_r2.tex   

\end{document}

%% file: list_of_authors_r2.tex
%
\author{                                                                      
V.M.~Abazov,$^{33}$                                                           
B.~Abbott,$^{70}$                                                             
M.~Abolins,$^{61}$                                                            
B.S.~Acharya,$^{27}$                                                          
M.~Adams,$^{48}$                                                              
T.~Adams,$^{46}$                                                              
M.~Agelou,$^{17}$                                                             
J.-L.~Agram,$^{18}$                                                           
S.H.~Ahn,$^{29}$                                                              
M.~Ahsan,$^{55}$
% ...
\\                                                                            
\vskip 0.30cm                                                             
\centerline{(D\O\ Collaboration)}                                             
\vskip 0.30cm                                                                 
}
\affiliation{                                                                 
\centerline{$^{1}$Universidad de Buenos Aires, Buenos Aires, Argentina}       
\centerline{$^{2}$LAFEX, Centro Brasileiro de Pesquisas F{\'\i}sicas,         
                  Rio de Janeiro, Brazil}                                     
\centerline{$^{3}$Universidade do Estado do Rio de Janeiro,                   
                  Rio de Janeiro, Brazil}                                     
\centerline{$^{4}$Instituto de F\'{\i}sica Te\'orica, Universidade            
                  Estadual Paulista, S\~ao Paulo, Brazil}                     
\centerline{$^{5}$Simon Fraser University, Burnaby, Canada, University of     
                  Alberta, Edmonton, Canada,}                                 
\centerline{McGill University, Montreal, Canada and York University,          
                  Toronto, Canada}                                            
\centerline{$^{6}$Institute of High Energy Physics, Beijing,                  
                  People's Republic of China}                                 
\centerline{$^{7}$Universidad de los Andes, Bogot\'{a}, Colombia}             
\centerline{$^{8}$Charles University, Center for Particle Physics,            
                  Prague, Czech Republic}                                     
\centerline{$^{9}$Czech Technical University, Prague, Czech Republic}         
\centerline{$^{10}$Institute of Physics, Academy of Sciences, Center          
                  for Particle Physics, Prague, Czech Republic}               
\centerline{$^{11}$Universidad San Francisco de Quito, Quito, Ecuador}        
\centerline{$^{12}$Laboratoire de Physique Corpusculaire, IN2P3-CNRS,         
                 Universit\'e Blaise Pascal, Clermont-Ferrand, France}        
\centerline{$^{13}$Laboratoire de Physique Subatomique et de Cosmologie,      
                  IN2P3-CNRS, Universite de Grenoble 1, Grenoble, France}     
\centerline{$^{14}$CPPM, IN2P3-CNRS, Universit\'e de la M\'editerran\'ee,     
                  Marseille, France}                                          
\centerline{$^{15}$Laboratoire de l'Acc\'el\'erateur Lin\'eaire,              
                  IN2P3-CNRS, Orsay, France}                                  
\centerline{$^{16}$LPNHE, Universit\'es Paris VI and VII, IN2P3-CNRS,         
                  Paris, France}                                              
\centerline{$^{17}$DAPNIA/Service de Physique des Particules, CEA, Saclay,    
                  France}                                                     
\centerline{$^{18}$IReS, IN2P3-CNRS, Universit\'e Louis Pasteur, Strasbourg,  
                  France and Universit\'e de Haute Alsace, Mulhouse, France}  
\centerline{$^{19}$Institut de Physique Nucl\'eaire de Lyon, IN2P3-CNRS,      
                   Universit\'e Claude Bernard, Villeurbanne, France}         
\centerline{$^{20}$RWTH Aachen, III. Physikalisches Institut A,               
                   Aachen, Germany}                                           
\centerline{$^{21}$Universit{\"a}t Bonn, Physikalisches Institut,             
                  Bonn, Germany}                                              
\centerline{$^{22}$Universit{\"a}t Freiburg, Physikalisches Institut,         
                  Freiburg, Germany}                                          
\centerline{$^{23}$Universit{\"a}t Mainz, Institut f{\"u}r Physik,            
                  Mainz, Germany}                                             
\centerline{$^{24}$Ludwig-Maximilians-Universit{\"a}t M{\"u}nchen,            
                   M{\"u}nchen, Germany}                                      
\centerline{$^{25}$Fachbereich Physik, University of Wuppertal,               
                   Wuppertal, Germany}                                        
\centerline{$^{26}$Panjab University, Chandigarh, India}                      
\centerline{$^{27}$Tata Institute of Fundamental Research, Mumbai, India}     
\centerline{$^{28}$University College Dublin, Dublin, Ireland}                
\centerline{$^{29}$Korea Detector Laboratory, Korea University,               
                   Seoul, Korea}                                              
\centerline{$^{30}$CINVESTAV, Mexico City, Mexico}                            
\centerline{$^{31}$FOM-Institute NIKHEF and University of                     
                  Amsterdam/NIKHEF, Amsterdam, The Netherlands}               
\centerline{$^{32}$University of Nijmegen/NIKHEF, Nijmegen, The               
                  Netherlands}                                                
\centerline{$^{33}$Joint Institute for Nuclear Research, Dubna, Russia}       
\centerline{$^{34}$Institute for Theoretical and Experimental Physics,        
                  Moscow, Russia}                                             
\centerline{$^{35}$Moscow State University, Moscow, Russia}                   
\centerline{$^{36}$Institute for High Energy Physics, Protvino, Russia}       
\centerline{$^{37}$Petersburg Nuclear Physics Institute,                      
                   St. Petersburg, Russia}                                    
\centerline{$^{38}$Lund University, Lund, Sweden, Royal Institute of          
                   Technology and Stockholm University, Stockholm,            
                   Sweden and}                                                
\centerline{Uppsala University, Uppsala, Sweden}                              
\centerline{$^{39}$Lancaster University, Lancaster, United Kingdom}           
\centerline{$^{40}$Imperial College, London, United Kingdom}                  
\centerline{$^{41}$University of Manchester, Manchester, United Kingdom}      
\centerline{$^{42}$University of Arizona, Tucson, Arizona 85721, USA}         
\centerline{$^{43}$Lawrence Berkeley National Laboratory and University of    
                  California, Berkeley, California 94720, USA}                
\centerline{$^{44}$California State University, Fresno, California 93740, USA}
\centerline{$^{45}$University of California, Riverside, California 92521, USA}
\centerline{$^{46}$Florida State University, Tallahassee, Florida 32306, USA} 
\centerline{$^{47}$Fermi National Accelerator Laboratory, Batavia,            
                   Illinois 60510, USA}                                       
\centerline{$^{48}$University of Illinois at Chicago, Chicago,                
                   Illinois 60607, USA}                                       
\centerline{$^{49}$Northern Illinois University, DeKalb, Illinois 60115, USA} 
\centerline{$^{50}$Northwestern University, Evanston, Illinois 60208, USA}    
\centerline{$^{51}$Indiana University, Bloomington, Indiana 47405, USA}       
\centerline{$^{52}$University of Notre Dame, Notre Dame, Indiana 46556, USA}  
\centerline{$^{53}$Iowa State University, Ames, Iowa 50011, USA}              
\centerline{$^{54}$University of Kansas, Lawrence, Kansas 66045, USA}         
\centerline{$^{55}$Kansas State University, Manhattan, Kansas 66506, USA}     
\centerline{$^{56}$Louisiana Tech University, Ruston, Louisiana 71272, USA}   
\centerline{$^{57}$University of Maryland, College Park, Maryland 20742, USA} 
\centerline{$^{58}$Boston University, Boston, Massachusetts 02215, USA}       
\centerline{$^{59}$Northeastern University, Boston, Massachusetts 02115, USA} 
\centerline{$^{60}$University of Michigan, Ann Arbor, Michigan 48109, USA}    
\centerline{$^{61}$Michigan State University, East Lansing, Michigan 48824,   
                   USA}                                                       
\centerline{$^{62}$University of Mississippi, University, Mississippi 38677,  
                   USA}                                                       
\centerline{$^{63}$University of Nebraska, Lincoln, Nebraska 68588, USA}      
\centerline{$^{64}$Princeton University, Princeton, New Jersey 08544, USA}    
\centerline{$^{65}$Columbia University, New York, New York 10027, USA}        
\centerline{$^{66}$University of Rochester, Rochester, New York 14627, USA}   
\centerline{$^{67}$State University of New York, Stony Brook,                 
                   New York 11794, USA}                                       
\centerline{$^{68}$Brookhaven National Laboratory, Upton, New York 11973, USA}
\centerline{$^{69}$Langston University, Langston, Oklahoma 73050, USA}        
\centerline{$^{70}$University of Oklahoma, Norman, Oklahoma 73019, USA}       
\centerline{$^{71}$Brown University, Providence, Rhode Island 02912, USA}     
\centerline{$^{72}$University of Texas, Arlington, Texas 76019, USA}          
\centerline{$^{73}$Southern Methodist University, Dallas, Texas 75275, USA}   
\centerline{$^{74}$Rice University, Houston, Texas 77005, USA}                
\centerline{$^{75}$University of Virginia, Charlottesville, Virginia 22901,   
                   USA}                                                       
\centerline{$^{76}$University of Washington, Seattle, Washington 98195, USA}  
}                                                                             

%% file: acknowledgement_paragraph_r2.tex
%
We thank the staffs at Fermilab and collaborating institutions, 
and acknowledge support from the 
Department of Energy and National Science Foundation (USA),  
Commissariat  \` a l'Energie Atomique and 
CNRS/Institut National de Physique Nucl\'eaire et 
de Physique des Particules (France), 
Ministry of Education and Science, Agency for Atomic 
   Energy and RF President Grants Program (Russia),
CAPES, CNPq, FAPERJ, FAPESP and FUNDUNESP (Brazil),
Departments of Atomic Energy and Science and Technology (India),
Colciencias (Colombia),
CONACyT (Mexico),
KRF (Korea),
CONICET and UBACyT (Argentina),
The Foundation for Fundamental Research on Matter (The Netherlands),
PPARC (United Kingdom),
Ministry of Education (Czech Republic),
Natural Sciences and Engineering Research Council and 
WestGrid Project (Canada),
BMBF and DFG (Germany),
A.P.~Sloan Foundation,
Research Corporation,
Texas Advanced Research Program,
and the Alexander von Humboldt Foundation.

%% file: paper_arx.bbl
\begin{thebibliography}{99}
\bibitem{charge}
Charge conjugate states are always implied in this Letter.
\end{thebibliography}
